\pdfoutput=1
\PassOptionsToPackage{square,comma,numbers,sort&compress,super}{natbib} 
\documentclass{article}
\usepackage[preprint]{nips_2018}
\usepackage{comment,multirow,floatrow}

\usepackage[utf8]{inputenc} 
\usepackage[T1]{fontenc}    
\newfloatcommand{capbtabbox}{table}[][\FBwidth]
\usepackage{soul}
\usepackage{url}
\usepackage[utf8]{inputenc}
\usepackage[font=small]{caption}
\usepackage{graphicx}
\usepackage{amsmath}
\usepackage{booktabs}
\usepackage{algorithm}
\usepackage{algorithmic}
\usepackage{xspace}
\usepackage{subfigure,xcolor,amsfonts}
\graphicspath{ {fig/} }
\usepackage{hyperref}
\usepackage{cleveref}
\crefname{section}{}{\S\S}
\crefdefaultlabelformat{\S#2#1#3}
\hypersetup{colorlinks={true},linkcolor={blue},citecolor=blue}
\usepackage{enumitem}
\setitemize{noitemsep,topsep=0pt,parsep=0pt,partopsep=0pt}
\newcommand{\R}{\mathbb{R}}
\newcommand{\xr}[1]{{\color{black} #1}}
\newcommand{\hx}[1]{{\color{purple} #1}}
\newcommand{\Emph}{\textbf}
\newcommand{\W}{$W \in \R^{n \times l}$, where $n$}
\newcommand{\WW}{$W \in [ W_1, W_2, ..., W_i ]^T, W_i \in \R^{n_i \times l_i}$}
\DeclareMathOperator*{\argmin}{arg\,min}

\newcommand{\sys}{Saec\xspace}

\title{\sys: 
Similarity-Aware Embedding Compression \\in Recommendation Systems}

\author{
Xiaorui Wu$^{1,2}$
\and
Hong Xu$^1$\and
Honglin Zhang$^2$\and
Huaming Chen$^2$\and
Jian Wang$^2$\\
$^1$Department of Computer Science, City University of Hong Kong\\
$^2$Platform and Content Group, Tencent\\
}

\begin{document}

\maketitle

\begin{abstract}



Production recommendation systems rely on embedding methods to represent
various features. 
An impeding challenge in practice is that the large embedding matrix incurs
substantial memory footprint in serving as
the number of features grows over time. 
We propose a similarity-aware embedding matrix compression method called \sys
to address this challenge.
\sys clusters similar features within a field to reduce the embedding matrix
size. 
\sys also adopts a fast clustering optimization based on feature frequency to
drastically improve clustering time. 
We implement and evaluate \sys on Numerous, the production distributed machine
learning system in Tencent, with 10-day worth of feature data
from QQ mobile browser.
Testbed experiments show that \sys reduces the number of embedding vectors by
two orders of magnitude, compresses the embedding size by $\sim$27x, and
delivers the same AUC and log loss performance.

\end{abstract}
\section{Introduction}
\label{sec:intro}

Embedding methods have attracted much attention recently. 
They are used in applications that need to learn
continuous representations from discrete features. 
Word embedding \cite{le2014distributed,marcus1993building} is
widely used in natural language processing (NLP), and embedding vector is
also adopted in knowledge graphs \cite{bordes2013translating}, social networks \cite{chen2016entity}, and recommendation systems \cite{huang2013learning,covington2016deep}. 
One can use an embedding vector to represent a
word in NLP or a feature in recommendation systems.

Although embedding methods show promising benefits, they pose serious
challenges especially when used in recommendation systems. Particularly, the
size of an embedding matrix grows linearly with the number of the symbols. 
In a recommendation system, the number of features (i.e. symbols) can be
hundreds of billions, which suggests that the size of the embedding matrix can
be up to hundreds of GBs \cite{numFeatures}. In order to provide efficient
lookup performance, embedding matrix has to be maintained in the DRAM.
Yet DRAM is expensive, especially the high-density modules
\cite{shortDRAM,ENGS18}, and its capacity is limited (typically 128GB).
For example a 32GB DDR4-2400 server DRAM module costs about USD\$300–\$350 as
of Feb 2019 \cite{mem_price}; having 128GB memory implies a per-server cost of
over USD\$1,200. Further, storing the embedding matrix on multiple servers
and using distributed serving leads to more complicated issues like
synchronization. 
Therefore it is crucial to minimize the memory footprint of 
embedding matrix for recommendation systems.

Model compression is widely used as an effective solution to reduce the number
of parameters and the model size.
Various methods have been proposed, such as parameter pruning 
\cite{han2015learning}, parameter sharing 
\cite{han2015deep,chen2015compressing,wu2018deep}, compact architecture 
\cite{wen2016learning,chen2018learning,shu2017compressing}, parameter
quantization \cite{arora2014provable}, and knowledge distillation 
\cite{hinton2015distilling}. Recently model compression has also been studied
for embedding matrix \cite{chen2018learning,shu2017compressing}.%

\begin{figure}
\centering     
\includegraphics[width=0.7\linewidth]{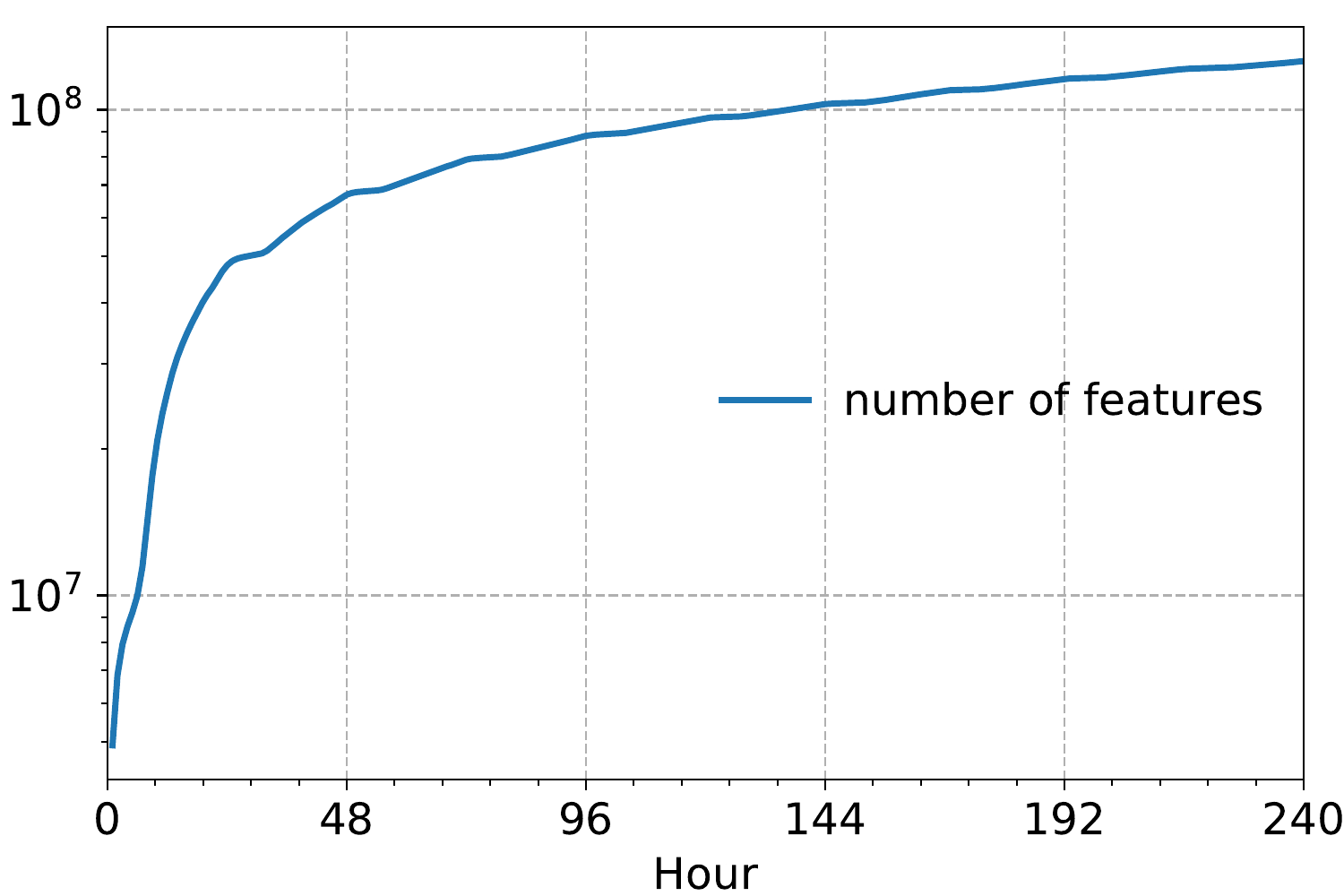}
\caption{
Time series of the number of features in a 10-day feature data
trace from Tencent's QQ mobile browser. Notice the
log scale for the y-axis. 
}
\label{fig:a}
\end{figure}

Unfortunately existing compression methods do not work well for
recommendation systems.
First, most methods focus on CNNs, particularly the fully-connected
layers which account for most of the model parameters \cite{han2015deep}. 
However, in an
recommendation model, fully-connected layers only occupy a small fraction
of the parameters \cite{huang2013learning,covington2016deep}.
Second, compression methods for embedding matrix 
\cite{chen2018learning,shu2017compressing} are designed for NLP
problems. 
The differences between NLP and recommendation systems are notable. 
In NLP, the number of words for the embedding matrix is fixed, whereas the number
of features for recommendation tasks in practice is growing constantly as new
data become available.
Figure~\ref{fig:a} 
shows that the number of features grows from less than $10^7$ to over $10^8$
in a 10-day production feature data trace from \xr{Tencent.} 
(see \cref{sec:setup} for more details).
Further, embedding vectors may have different lengths in recommendation systems
instead of a uniform length in NLP. In the embedding matrix from the same 
{production} feature data trace,
88.95\% vectors have a length of 9 and 11.05\% have 17.

In this paper we propose to utilize two unique characteristics of
recommendation systems,
feature similarity and feature frequency, for embedding compression. 
First, embedding representations are usually over-parameterized: similar
features have similar meanings and effects. We can use
one embedding vector to
express multiple similar features and in turn reduce the size of embedding
matrix.
Second, features appear in training data with different
frequencies. Consequently the rare or less frequent features receive little
training and have little impact or benefit to the overall performance \cite{covington2016deep}. 
We can also compress these features with similar ones to reduce model
size and improve their aggregated training effect.

We design an embedding compression framework called \sys to mine the 
similarity across features in recommendation systems. We make three new contributions.
\begin{itemize}
	\item  We propose a novel compression method based on $k$-means for
	embedding matrix in recommendation systems. Our method is field-aware: we
	apply
	$k$-means clustering
	on each {\em field} of features instead of on all features. A {field}
	contains a group of similar features with the same length generated by
similar inputs in feature engineering. Field-aware
clustering thus overcomes the challenge of non-uniform length of embedding
vectors in recommendation systems.
	\item We design a fast clustering method to make \sys practical. As the
	feature data are enormous, clustering every single feature
	has formidable computation costs and is infeasible in practice
	where the model is trained say hourly. 
	Since the feature frequency follows a heavy-tailed distribution, we cluster
	only a small fraction of the most frequent features and assign the
	other features to their closet centroids.
	\item We implement \sys in the production deep learning system in 
	Tencent and evaluate it in a large-scale testbed with the production
	feature data trace from QQ mobile browser. 
	Our evaluation shows that \sys (1) {delivers the same
	AUC and log loss performance with $\sim$27x	effective compression ratio
	taking into account its own overhead}; and (2)
	accelerates	clustering by 32x with fast clustering optimization.

\end{itemize}

We are optimizing our implementation of \sys and driving for the full deployment
in Tencent.




\section{Design}
\label{sec:design_clustering}
We now present \sys, our similarity-aware embedding compression
framework based on $k$-means clustering. 

\subsection{Overview}
\label{sec:design_overview}

\begin{figure}[ht]
\centering
\includegraphics[width=0.8\linewidth]{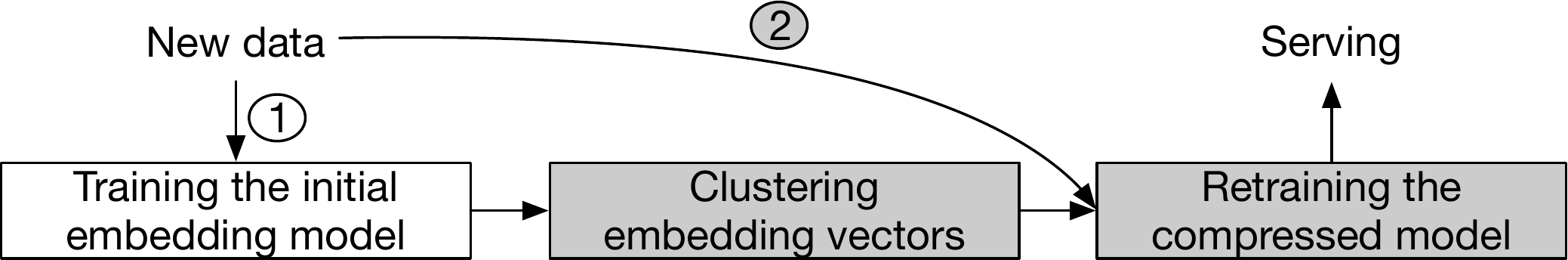}
\caption{Training process in \sys. The gray steps are new in \sys. }
\label{fig:framework}
\end{figure}
\sys's training process is shown in Figure~\ref{fig:framework}. 
Like existing recommendation systems, \sys trains the embedding model
continuously in a fixed time internal of say one hour with newly generated data.
Each hour's data are processed just once in one epoch instead of
multiple epochs as used for other deep learning models in general. 
This initial model can be enormous.
\sys then applies clustering to compress it into a much
smaller model with only a subset of the original embedding vectors.
Directly applying the compressed model generally results in poor prediction
performance \cite{han2015deep}.
Thus \sys finally retrains the compressed model with the same data again in one
epoch to compensate for the performance loss of compression. 

As \sys does not change the training process for the initial model, we
focus on its clustering method (\cref{sec:clustering}) and retraining strategy 
(\cref{sec:retraining}) in the following. We also discuss an optimization to
accelerate clustering at the end (\cref{sec:design_fast}).

\subsection{Clustering Embedding Vectors}
\label{sec:clustering}

Our clustering method works on a field basis in order to overcome the variable
embedding vector length problem as mentioned in \cref{sec:intro}.
Our subsequent discussion assumes a particular field without loss of generality.
\vspace{-0.05cm}
\subsubsection{Field-Aware Clustering}
\label{sec:clustering_basics}

Each field has its own embedding matrix $W \in \mathbb{R}^
{n\times l}$ from the initial training, where $n$ is the number of
features and $l$ the length of embedding vectors.
In other words the embedding matrix is just a stack of embedding vectors, each
representing a unique feature.
\sys aims to cluster the original $n$ embedding vectors into just $k$
clusters, and for each cluster use one embedding vector to represent all the
rest. 
Thus after compression there are $k$ representative embedding vectors which
constitute the field's
{\em codebook} $C$. That is, $C = \left\{C_1, C_2, ...,
C_k\right\}$ where $C_i$ is an $l$-dimensional vector.

\begin{figure}
\centering     
\subfigure[]{
\label{fig:lookup_old}\includegraphics
[width=0.485\linewidth]{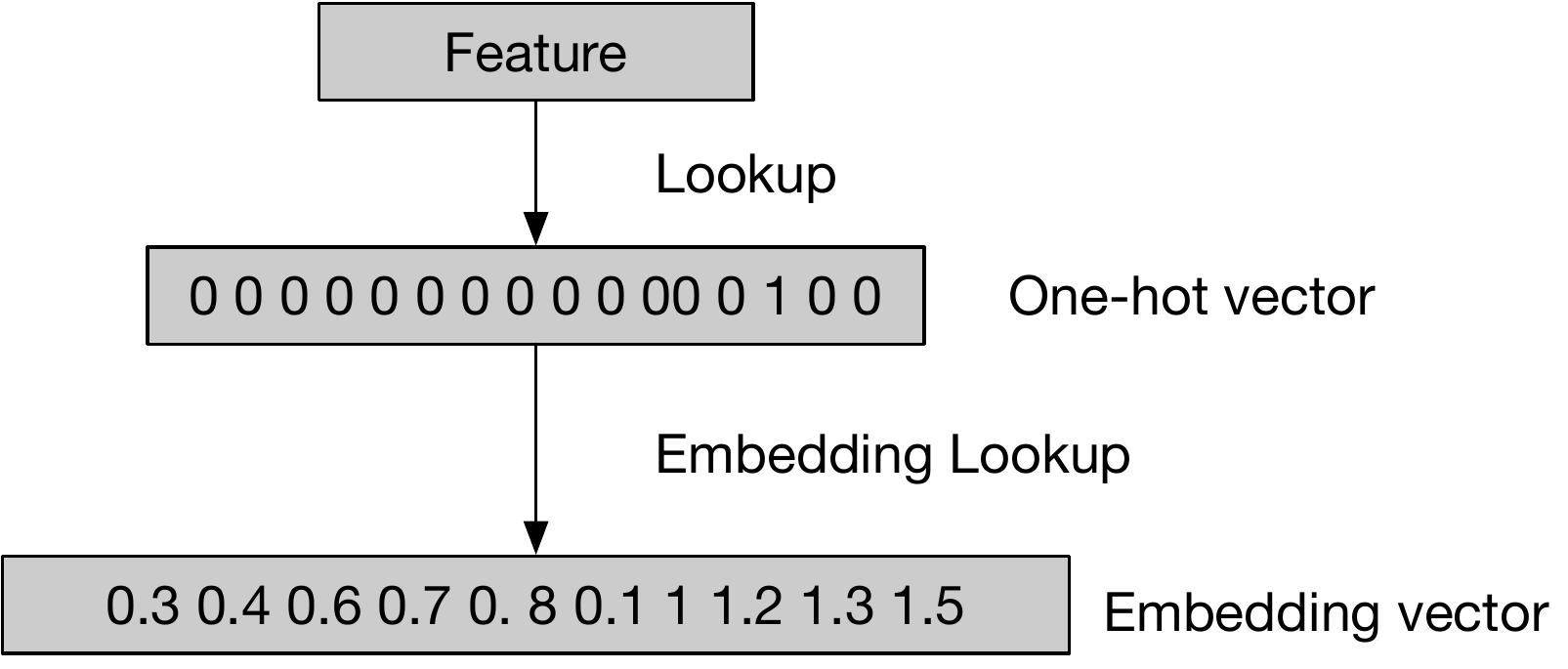}}
\subfigure[]{
\label{fig:lookup_new}\includegraphics
[width=0.485\linewidth]{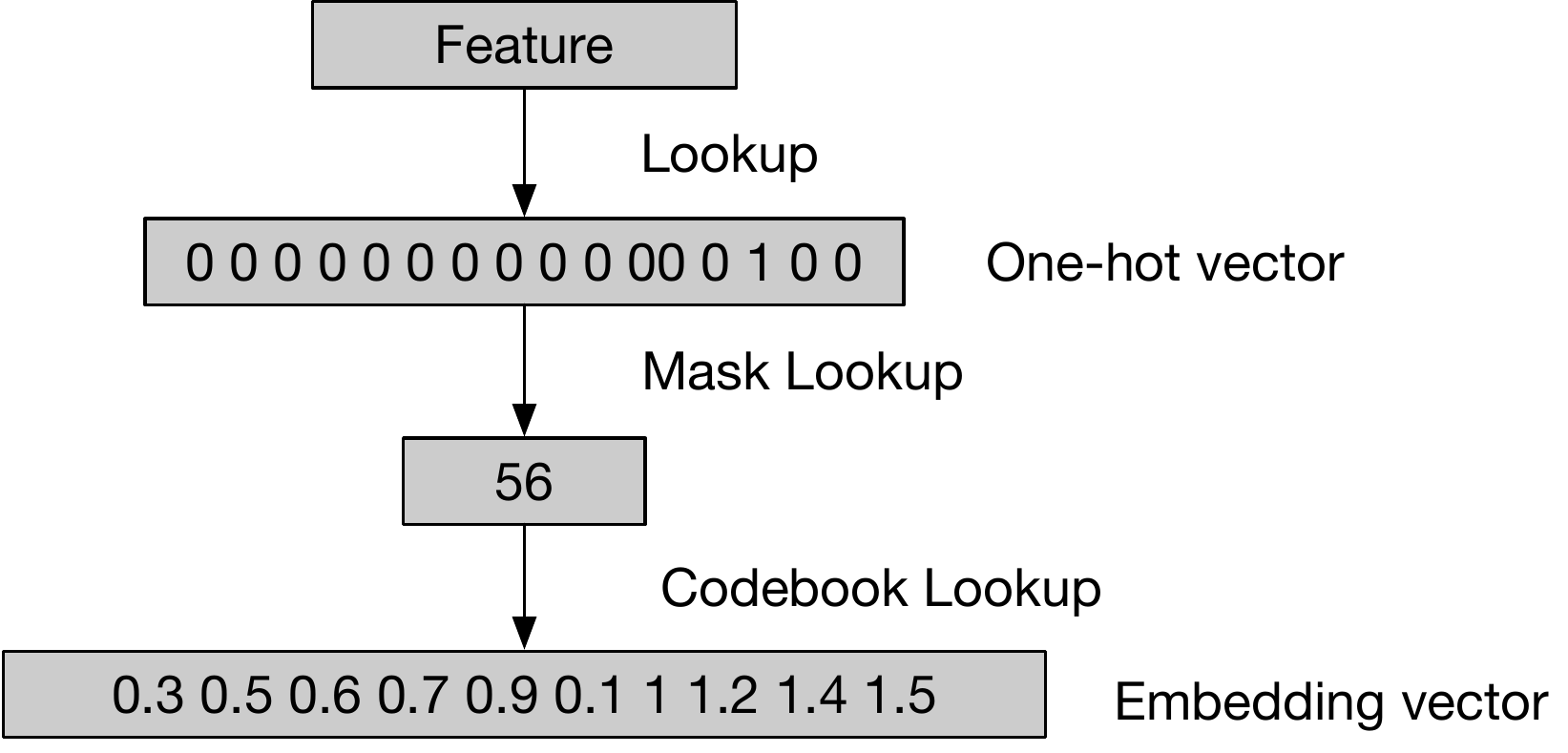}}
\caption{Embedding vector retrieval in \sys for (a) uncompressed features and 
(b) for compressed features.}
\label{fig:retrieval}
\end{figure}

Now without clustering, we generally use a one-hot vector $X$ to encode a
feature, and retrieve its corresponding embedding vector by a lookup with
the embedding matrix $W$: $V = W^T X$ as shown in Figure~\ref{fig:lookup_old}.
With clustering, to retrieve an embedding vector from the codebook, \sys
in addition associates each feature with a {\em mask}. 
A table lookup transforms the feature's one-hot vector to its mask which is an
index that points to the corresponding representative embedding vector in the
codebook. Figure~\ref{fig:lookup_new} depicts the process.
Generally $k \ll n$, and the codebooks and masks that \sys needs to store after
clustering are also small.

\begin{figure*}
\centering 
\subfigure[CDF of per-field number of features]{
\label{fig:slot_features}
\includegraphics[width=0.33\textwidth]{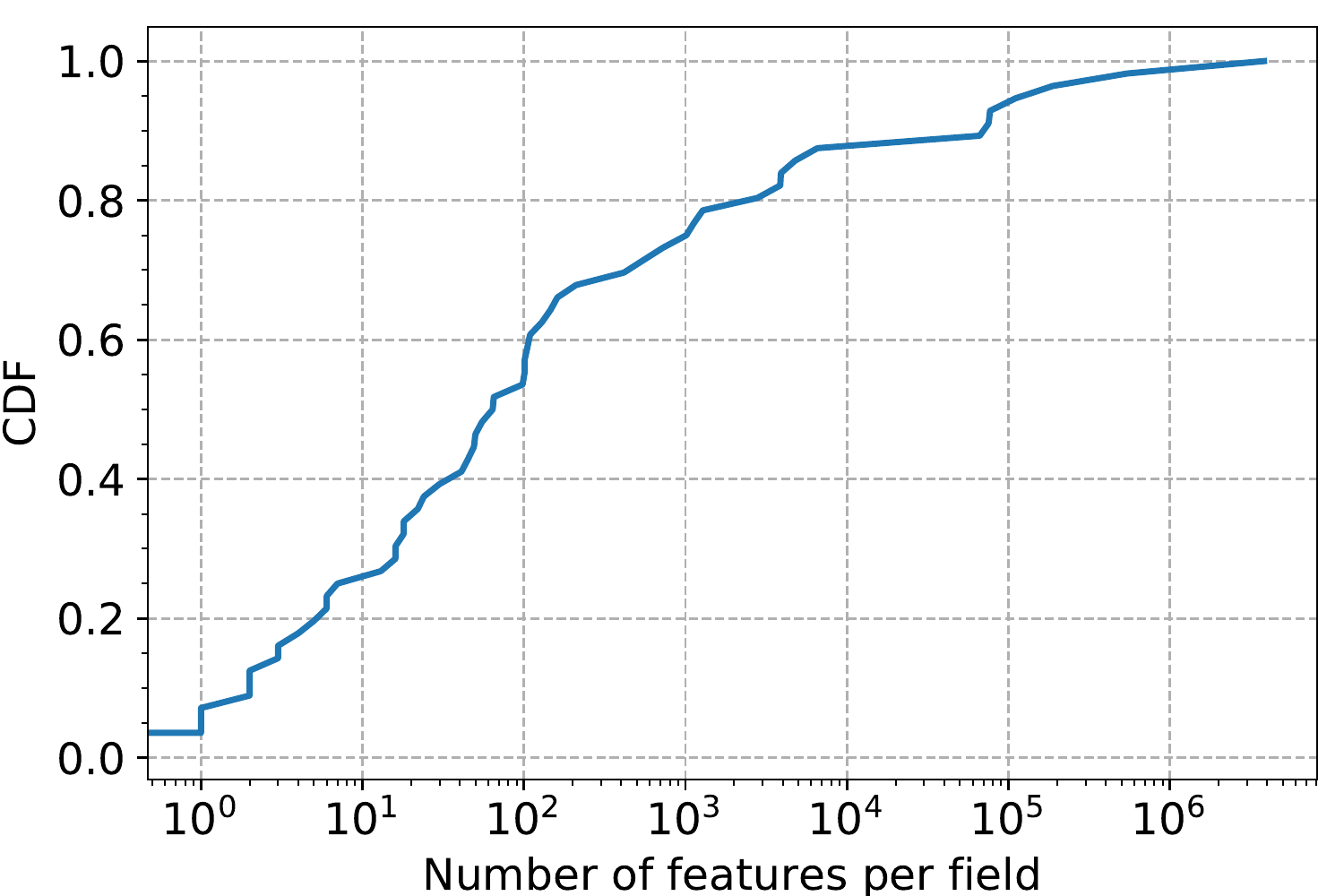}}%
\subfigure[CDF of feature frequency]{
\label{fig:freq}
\includegraphics[width=0.33\textwidth]{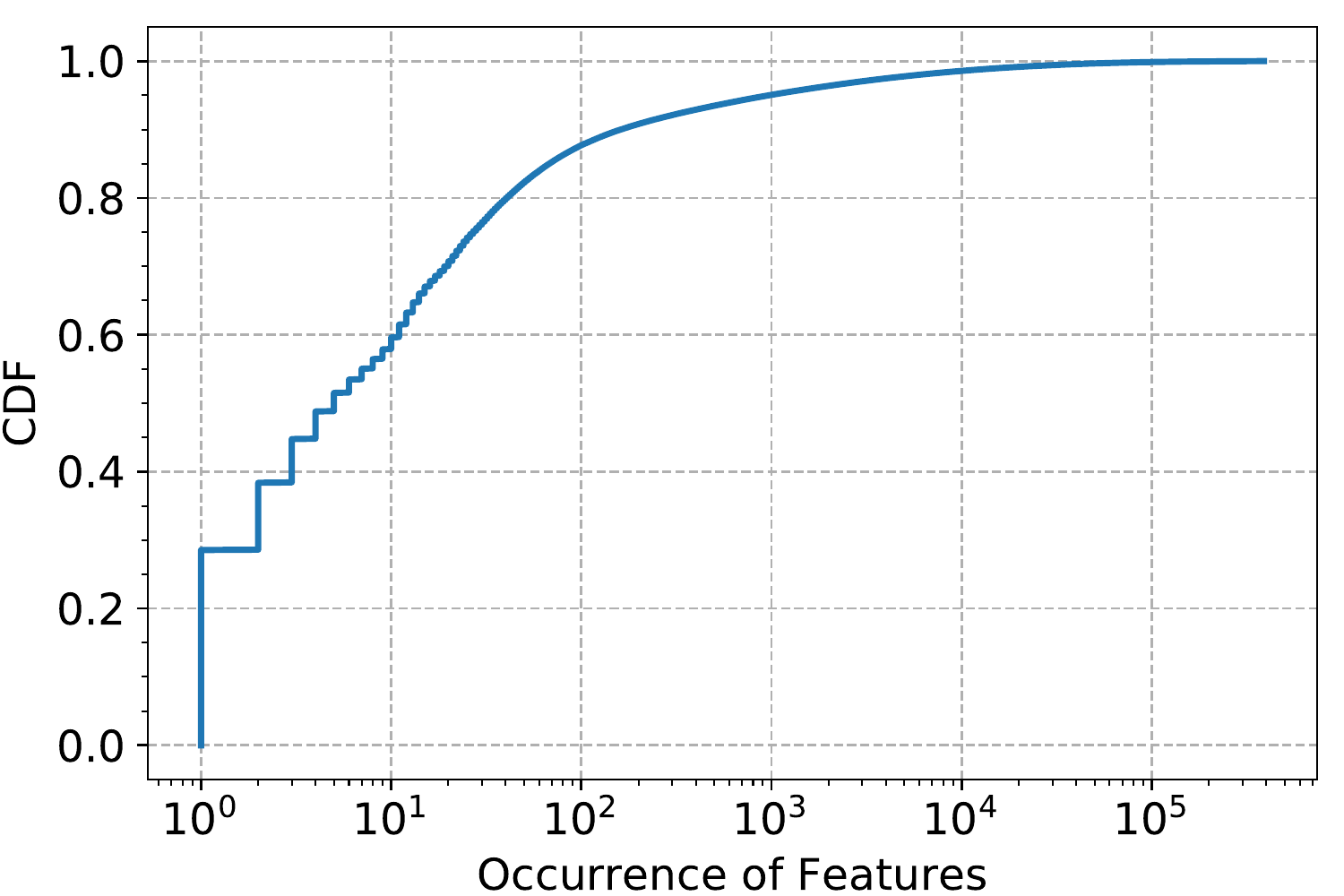}}%
\subfigure[\sys's clustering time]
{\label{fig:clustering_time}
\includegraphics[width=0.33\textwidth]{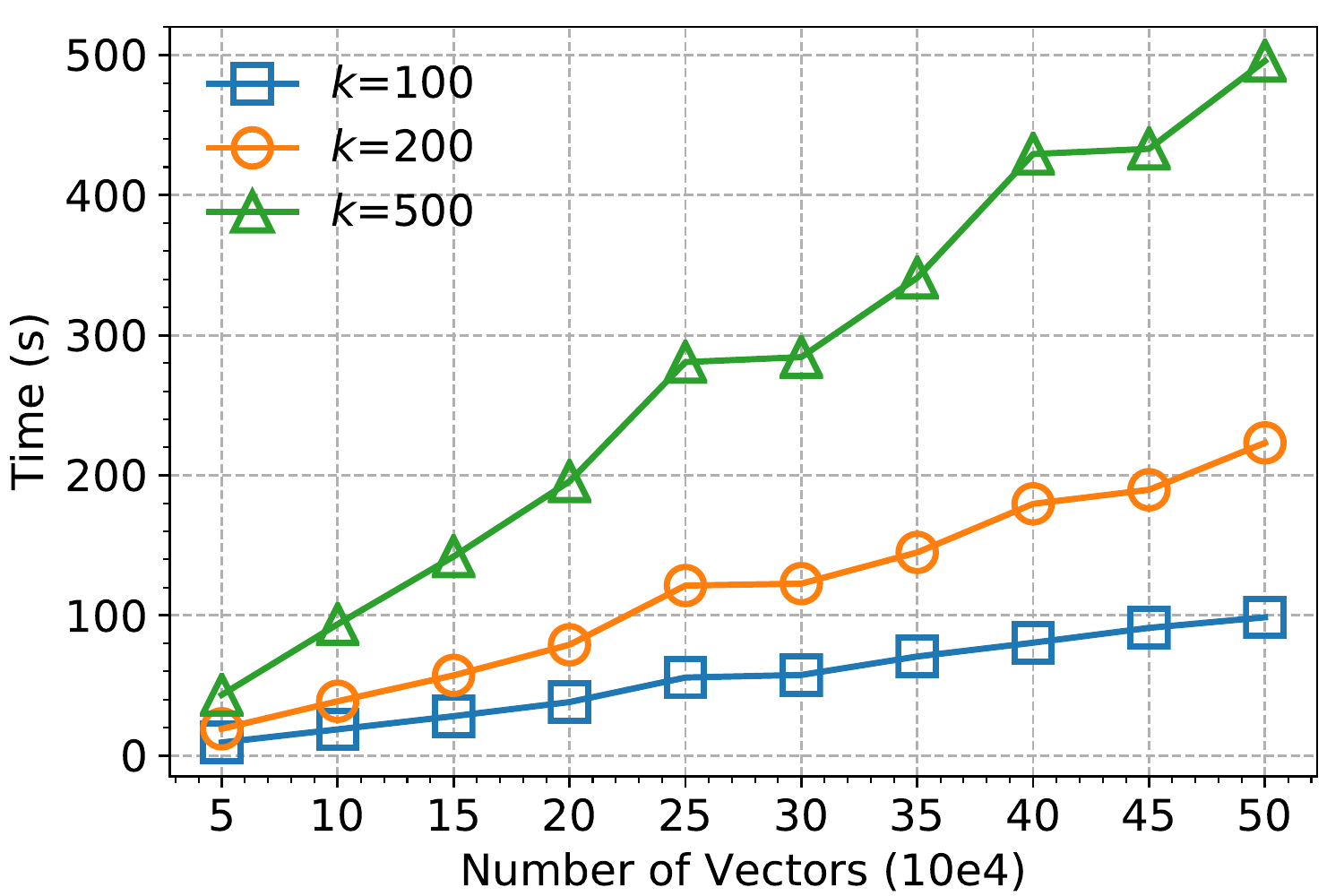}}%
\caption{Motivation figures for \sys's design. For (a) and (b) the 
features are from one hour of 10-day feature data trace from Tencent QQ
mobile browser. For (c) we show the time to run $k$-means clustering with
different
numbers of vectors using 1 CPU core of an Intel Xeon E5-2520 v3 at 2.4Ghz.}
\end{figure*}

The central task is then to design an efficient clustering method that yields
a good codebook.
Mathematically, the problem can be posed as an optimization to find the
best $k$ centroids, i.e. representative embedding vectors, that minimize the
Euclidean distance between the
original embedding vectors and the centroids:
\[
 \argmin_{C} \displaystyle\sum_{i=1}^{k} \displaystyle\sum_{j=1}^{n}
 \| W_j - C_i\|^2.
\]
This problem can then be solved by the well-known $k$-means clustering.

The number of features in a field is not uniform in practical recommendation
system. For instance Figure~\ref{fig:slot_features} shows the distribution of
per-field number of features in one hour of our 10-day feature data trace. 
We observe that only around 7\%
fields have more than
$10^5$ features; most have less than $10^4$ features. 
It thus makes sense to focus on clustering the large fields to ensure the
effectiveness of compression. 
We choose fields whose number of features is at least 100 times more than
$k$ to go through the clustering process. As a reference $k$ is on the order of
hundreds in our current design.

\vspace{-0.05cm}
\subsubsection{Centroid Initialization}
\label{sec:design_init}
Centroid initialization has a direct impact on the quality of clustering and
thus the model performance. Here we examine three initialization methods,
Random, $k$-means++, and Top-$k$ initialization which we propose.

\Emph{Random.} This method randomly chooses $k$ embedding vectors from the
embedding matrix as the initial centroids. It does not consider the
feature distribution or the parameter in the embedding vectors.

\Emph{$k$-means++.} This method chooses the first initial centroid uniformly at
random. After that, each subsequent centroid is chosen according to a weighted
probability based on the Euclidean distance between the potential
candidate and the existing centroids \cite{arthur2007k}. 
It aims to reduce the probability of having similar centroids.

\Emph{Top-$k$.} We propose this method to explore the feature
frequency characteristics of recommendation systems.
Figure~\ref{fig:freq} plots the CDF of feature frequency in our feature 
data trace. 
There is a clear heavy-tailed pattern: a very small number of frequent features
appear much more often in the training data and receive much better training. 
Top-$k$ simply chooses $k$ most frequent embedding vectors as the initial
centroids.

We conduct experiments in \cref{sec:result_init} to compare the
prediction performance of these methods. 
The results show that Top-$k$ and $k$-means++ are both viable choices and
$k$-means++
has the best accuracy performance. 
We thus choose $k$-means++ as the initialization method in the design.

\subsection{Retraining to Optimize Codebooks}
\label{sec:retraining}

Model compression inherently leads to loss in 
inference accuracy which has been reported for parameter
pruning, parameter sharing, etc \cite{han2015deep,han2015learning}.  
Retraining the compressed model is a common method to compensate the performance
loss \cite{amc}.

\sys also retrains the representative embedding vectors after clustering as the
final step.
Different from the initial training, in retraining the embedding
vectors have to be retrieved and updated indirectly in both the forward pass  
and backpropagation.
In the forward pass, \sys uses a feature's mask to retrieve its
corresponding representative embedding vector and then execute the forward
process. The rest of computation is the same as the initial training.
In backpropagation, each feature gets its own gradient. \sys then collects the
gradients from features in the same cluster according to their mask and average
them as the gradient of the representative embedding vector for updates.
In other words, the representative vectors receive more extensive training now.

\subsection{Fast Clustering} 
\label{sec:design_fast}

We have introduced \sys's clustering method so far. 
An issue of our design is that clustering entails excessive computation cost: for
production-scale systems, processing billions of embedding vectors may take
well over an hour which is the typical training interval for recommendation
models in practice.
Figure~\ref{fig:clustering_time} shows that running 
vanilla $k$-means over 50,000 vectors on a single thread with an Intel
Xeon CPU already takes about 500s with $k=500$.

To make \sys more practical, we introduce a fast clustering optimization. 
We only process a part (100$k$ in our design) of the most frequent
features and their embedding vectors in each field for clustering.
The other embedding vectors are simply assigned with the nearest cluster
centroids as their representative embedding vectors.

The rationale behind this design is that the clustering time grows
proportionally with the number of vectors processed; 
{Figure~\ref{fig:clustering_time} clearly shows this trend when we run
vanilla $k$-means.}
Yet as we showed in {Figure~\ref{fig:freq}}, many features appear with low
frequency. 
Prior work \cite{covington2016deep} shows that low frequency features cannot
be trained well and do not have much benefit to the model performance.
Reducing the number of features for clustering can therefore reduce the overhead
dramatically without degrading performance.
\section{Evaluation}
\label{sec:eva}

We present our evaluation results in this section.

\subsection{Setup}
\label{sec:setup}

\noindent{\bf Prototype.}
We implement \sys on Numerous, {a distributed machine learning system based on parameter server in Tencent.
Our prototype consists of $\sim$1k LoC in C++ and follows our design in 
\cref{sec:design_clustering} including the fast clustering optimization.
For load balance features are assigned to server nodes via hashing. 


\noindent{\bf Testbed.}
We use a container based CPU testbed. Throughout the experiments the embedding
models are trained with 50 worker containers and 40 server containers. Each
worker has 6 vCPUs and 15GB memory; each server container has 2 vCPUs and
12.8GB memory. 

\noindent{\bf Model and Datasets.}
We use Tencent's recommendation model and the embedding matrix
that serve the QQ mobile browser. The model is a modified DSSM 
\cite{denil2013predicting}. 
Different from the original
DSSM, we directly adopt the embedding method to train the embedding matrix from
scratch instead of using word2vec. Our model contains the embedding matrix and
multiple fully-connected layers. The size of fully-connected layers is also different from the original DSSM. 
\sys does not change these fully-connected layers.

We collect a 10-day feature data trace from the QQ mobile browser,
including
features about users, user activities (search query tokens, watch times, etc.),
news items, etc. as the
training dataset. 
The feature data trace, with tens of TB of data, is collected on
a hourly basis and has 240 segments each used for one epoch of training,
following
the same strategy of training in production. 

We compare \sys against Baseline which is the original model with uncompressed
embedding matrix. Both Baseline and \sys training use the same hyper-parameters:
we adopt Adagrad as the optimizer and use a learning rate of 0.001.

\noindent{\bf Performance Metrics.}
Recommendation systems are commonly evaluated using AUC and log loss. 

\begin{itemize}
\item {\bf AUC}, area under the receiver operating curve, reflects 
the probability that a uniformly drawn positive item is ranked higher than a
negative item by the model. It is the most widely used performance metric for
recommendation tasks.
AUC is calculated by:
\[
AUC = \int\limits_0^1 TPR(t)d_{FPR(t)}.
\]
Here $t$ is the threshold which decides the TPR (true positive rate) and FPR 
(false positive rate). AUC is a value which ignores the distribution of samples.
Generally, the higher the AUC is, the better the model is.

\item{\bf Log loss} is a natural measure of how close the predicted click
probability is to the true 0-1 label in a statistical view \cite{bishop2006pattern}. For
binary
classification tasks such as ours, the log loss is defined as:
\[
 -\frac{1}{n}\sum_{i=1}^n  y_i\log(p_i) + (1-y_i)\log(1-p_i),
\]
where $n$ is number of samples, $y_i$ is the true binary label of whether the
user clicks the item for sample $i$, and $p_i$ is the click probability 
given by the model.
The log loss value trends toward zero as the prediction becomes more
accurate. If the distribution of positive and negative samples are uniform, the
log loss is 0.693 for random prediction. 

\end{itemize}



\subsection{Overall Performance}
\label{sec:dssm}

\begin{table*}[!h]
\begin{center}
\resizebox{\textwidth}{!}{%
\begin{tabular}{|c|c|c|c|c|c|c|}
\hline
Day & \# vectors (Before)   & \# vectors (After) & AUC (Before) & AUC 
(After) & Log loss (Before) & Log loss (After) \\ \hline
1     & $4.78 \times 10^7$ & $5.91 \times 10^5$          & 0.737362    & 0.782574    & 0.402417         & 0.378207        \\ \hline
2     & $6.69 \times 10^7$  & $6.58 \times 10^5$          & 0.780317     & 0.800837    & 0.375378         & 0.363054         \\ \hline
3     & $7.90 \times 10^7$ & $7.38 \times 10^5$          & 0.798804     & 0.811197    & 0.358956          & 0.351107        \\ \hline
4     & $8.84 \times 10^7$ & $7.98 \times 10^5$          & 0.805723     & 0.81675   & 0.356759         & 0.349559        \\ \hline
5     & $9.62 \times 10^7$ & $8.50 \times 10^5$          & 0.80804   & 0.818024    & 0.355246         & 0.348736        \\ \hline
6     & $1.0 \times 10^8$ & $8.95 \times 10^5$          & 0.811398     & 0.81524    & 0.354769         & 0.349977         \\ \hline
7     & $1.07 \times 10^8$ & $9.39 \times 10^5$          & 0.815689    & 0.819183    & 0.345094          & 0.347014        \\ \hline
8     & $1.14 \times 10^8$ & $9.72 \times 10^5$          & 0.814011     & 0.819918    & 0.352489         & 0.343376        \\ \hline
9     & $1.20 \times 10^8$ & $1.0 \times 10^6$          & 0.816787     & 0.823498    & 0.308586         & 0.315366        \\ \hline
10    & $1.24 \times 10^8$ & $1.04 \times 10^6$          & 0.817091    & 0.822134    & 0.347485         & 0.342201        \\ \hline
\end{tabular}
}
\caption{Performance results with the embedding matrix at the last hour of each
day with the 10-day feature data trace. } \label{tab:timestamp}
\end{center}
\end{table*}

We first evaluate the overall prediction performance of the model with \sys. 
The data segments as mentioned before are continuously fed to the model. 
We set $k$ to
100 so the condition to trigger compression for a field is it has at least
$10^4$ features as explained in \cref{sec:clustering_basics}.

The embedding matrix keeps changing as the training proceeds with new feature
data.
Table~\ref{tab:timestamp} shows the performance results on a daily basis,
where the ten models trained after the last hour of each day are used. We make
several important observations.

First, as the number of embedding vectors in the original embedding matrix
grows steadily with time, \sys consistently achieves significant
compression by reducing two orders of magnitude for the number of embedding
vectors. 
Note \sys also introduces overheads with masks as discussed in 
\cref{sec:clustering_basics}.
Since $k$ is 100, we can encode the masks with 8 bits for each field. 
Now we can the obtain actual compression ratio of \sys, for say the final model
after
240 epochs which is the last row in Table~\ref{tab:timestamp}. 
Assume the length of vectors is 9 for simplicity.
Then the size of the original embedding matrix is $1.24 \times 10^8
\times 36$ bytes, while the size after compression is $1.24 \times 10^8 \times 1
+ 1.04 \times 10^6 \times 36 $ bytes.
That is, \sys achieves a compression ratio of 27.7.


Second, \sys's compression does not degrade performance; in fact in most cases,
the compressed model slightly outperforms the original one in both AUC and log
loss as shown in Table~\ref{tab:timestamp}.
The final AUC of \sys is 82.22\% which is 0.5\% higher than Baseline; the final
log loss of \sys is 0.342 while that of Baseline is 0.347.
We believe there are two reasons for the performance improvement  after
compression. 
For one, the low frequency features may not receive enough training, as a result
the original model has a good probability to give a wrong prediction with these 
poorly trained embedding vectors. 
After compression, more frequent embedding vectors are assigned to
the low frequency features and then retrained, which improves
the probability to make the right prediction.
In addition, the input feature data have noise which may affect the
accuracy of prediction. 
On the contrary, with \sys multiple features share
the same embedding vector and more data are actually used to
train these representative embedding vectors, which {alleviates the influence
of noise and leads to better prediction.

\begin{figure}[ht]
\centering
\includegraphics[width=0.7\linewidth]{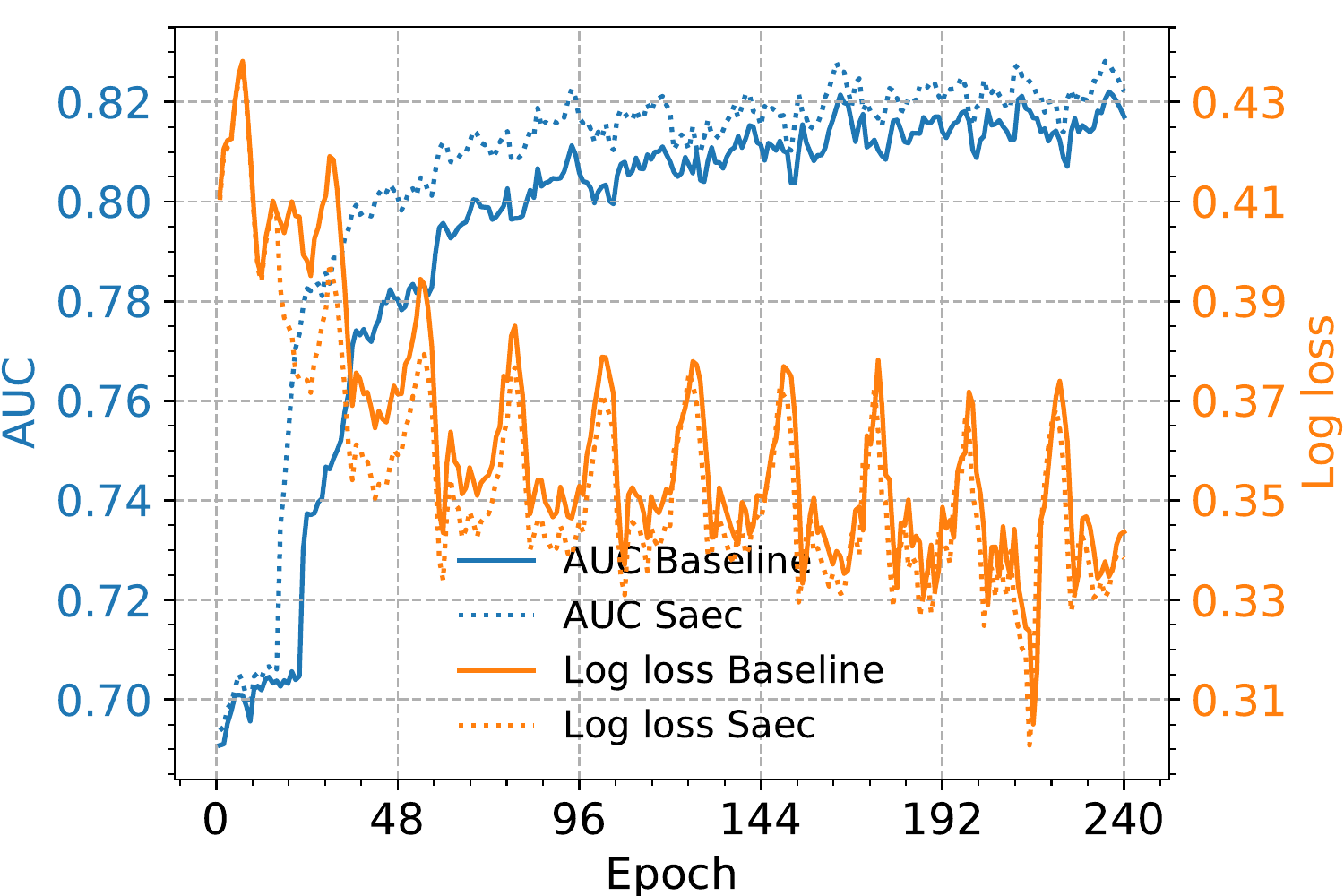}
\caption{Model performance in Baseline and \sys during the complete training
process.}
\label{fig:perf_240}
\end{figure}

Third, we also observe that while the AUC and log loss performance improves 
in general as time goes with more training, they
sometimes deteriorate (e.g log loss in day 8 is worse than day 7). 
This performance oscillation is normal during training. 
To see this, we plot the time series of AUC and log loss of both Baseline and
\sys for the 240-epoch training process in Figure~\ref{fig:perf_240}.
Clearly both performance metrics exhibit oscillation while they improve in
general as training proceeds. The log loss performance shows more volatile
oscillation than AUC. 
Nevertheless, \sys's curves closely track those of Baseline.



\begin{figure}[ht]
\centering
\includegraphics[width=0.7\linewidth]{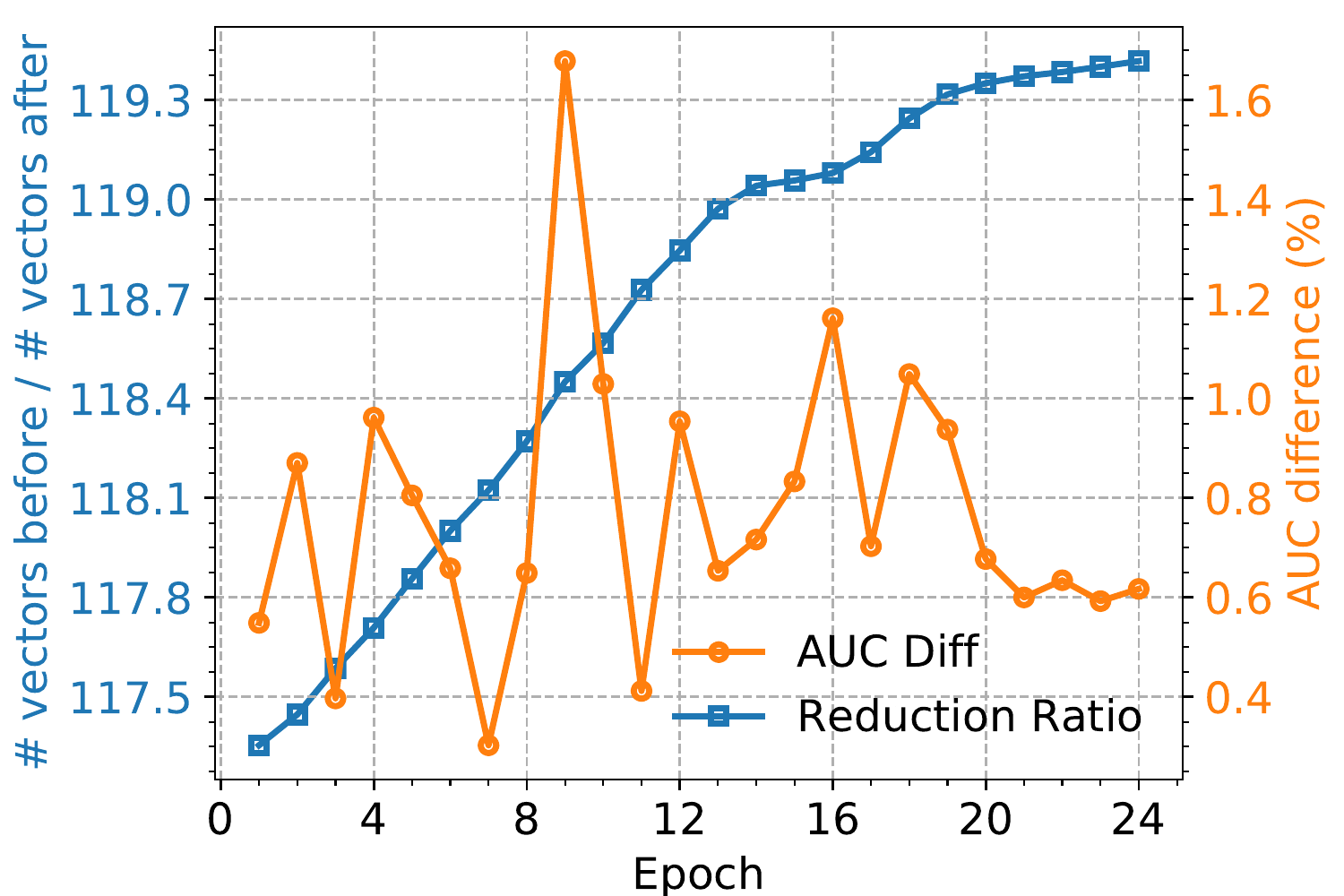}
\caption{\sys's reduction ratio in number of embedding vectors (\#
before / \# after) and AUC
improvement  (AUC after $-$ AUC before) in the last 24 epochs.}
\label{fig:hours}
\end{figure}

Lastly, we zoom into \sys's hourly performance within a day to better analyze
it.  
We plot \sys's reduction ratio in number of embedding vectors against its
AUC improvement (AUC after $-$ AUC before) for the last 24 epochs of training in
Figure \ref{fig:hours}. 
We observe that the reduction ratio steadily increases. 
This is because as the original embedding matrix gets larger more fields
meet \sys's clustering threshold and undergo compression. 
At the same time, the AUC of \sys is always better than Baseline which is
consistent with the results presented before.

\subsection{Effect of \sys's Design Choices}
\label{sec:result_init}

We now study the effect of \sys's design choices on performance.
First we look at the centroid initialization methods.  
We disable our fast clustering optimization to better compare the
clustering time of different methods.
{We only show the results at the 24-epoch as an example since clustering
takes much time without the fast optimization.}
Clustering time here measures the time it
takes for all 40 server
containers (80 vCPU cores) to finish clustering all eligible fields and
generating the codebooks
and masks. Note workers do not perform clustering.

Table~\ref{tab:dif_init} shows the results of three initialization methods
mentioned in \cref{sec:design_init}: Random, $k$-means++, and Top-$k$ which we
propose.
All methods achieve similar AUC of $\sim$77\% and log loss of $\sim$0.38. 
Among them $k$-means++ achieves the best performance in both AUC and
log loss.
$k$-means++ works best since it chooses more different embedding vectors than
the existing centroids
and reduces the probability of having similar centroids.
In terms of clustering time, our Top-$k$ method performs the best and is 5.9\%
faster than
$k$-means++ and 9.1\% faster than Random.
We believe both $k$-means++ and Top-$k$ are viable design choices.
Our current design adopts $k$-means++ to obtain the best prediction performance.

\begin{table}[h]
\begin{center}
\resizebox{0.6\linewidth}{!}{%
\begin{tabular}{c|ccc}
\hline
Method   & AUC      & Log loss    & Clustering time (s) \\ \hline
Random   & 0.775876 & 0.381991 & 3718     \\ \hline
$k$-means++ & 0.779863 & 0.37972  & 3594     \\ \hline
Top-$k$     & 0.77503 & 0.382567 & 3381     \\ \hline
\hline
Fast clustering & {0.782574} & {0.378207} & {111.54}      \\
\hline
\end{tabular}
}
\caption{Results of different centroid initialization methods and our fast
clustering design. Each setting uses the same $k$ of 100. Clustering time
is measured for all 40 server containers (80 vCPU cores) to finish clustering
the embedding matrix.} 
\label{tab:dif_init}
\end{center}
\end{table}

Next we investigate the effectiveness of fast clustering, another key design
choice in \cref{sec:design_fast}.
The last row of Table~\ref{tab:dif_init} shows its results for the same
embedding matrix with $k$-means++ for centroid initialization.
By working over only the 100$k$ most frequent embedding vectors of a field,
fast clustering
significantly reduces the clustering time by over
32x to less than 2 minutes compared to vanilla $k$-means++. 
The performance with fast clustering is slightly better due to better training
for the representative embedding vectors as we explained before.
The results thus demonstrate the effectiveness of our fast clustering design.

\subsection{Effect of $k$}

In this section we look at the effect of number of clusters $k$ on \sys.
We use various values of $k$ and show \sys's performance at the 24th epoch as
an example.
Table~\ref{tab:dif_size} shows the results. 
Note that as $k$ increases the reduction ratio
in number of embedding vectors naturally becomes smaller since there are more
representative vectors after compression and fewer fields can be compressed.
We thus do not show the reduction ratio in Table~\ref{tab:dif_size}.

First, observe that the clustering cost increases with $k$: for $k=1000$ the
clustering time is
$\sim$41x
longer than $k=100$, and far exceeds one hour (with fast clustering).
This is because though fewer fields are eligible for clustering with a larger
$k$, \sys requires
more embedding vectors to undergo processing since the most frequent 100$k$
vectors are used in fast clustering.


\begin{table}[h]
\resizebox{0.5\textwidth}{!}{%
\begin{tabular}{c|ccc}
\hline
$k$      & AUC (\%)  & Log loss & Clustering Time (s) \\ \hline
{50}      & 77.9 & 0.380    & 67.27             \\ \hline
100      & 78.3 & 0.378    & 111.54              \\ \hline
200      & 78.3 & 0.378    & 228.65              \\ \hline
500      & 77.9 & 0.380    & 591.49              \\ \hline
1000     & 77.5 & 0.383    & 4608                \\ \hline
\hline
Baseline & 73.7 & 0.402    & $\sim$              \\ \hline
\end{tabular}
}
\caption{Results with varying values of $k$ in \sys at the 24-th epoch.}
\label{tab:dif_size}
\end{table}

Performance-wise, a larger $k$ beyond 200 leads to slightly worse results.
The AUC with $k=500$ and $k=1000$ is 0.3\% and 0.7\% lower than
with $k=100$, respectively. 
We believe this is caused by the large differences in the number of embedding
vectors undergoing clustering.
With a large $k$ more poorly-trained less-frequent embedding vectors may be
chosen as the initial centroids and influence the result.
Moreover, each representative embedding vector has less data in retraining with
a larger $k$. 
Lastly, we observe that a very small value of $k=50$ also leads to
inferior performance because the
compression becomes too coarse-grained and lose too much information.

Thus the results here justify our choice of using $k=100$ to achieve a good
tradeoff between compression and performance in \sys.

\section{Related Work}

We survey the most related work to \sys now.

The problem of large embedding matrix and its memory footprint has recently
received much attention \cite{ENGS18}.
The idea of using more efficient coding schemes for embedding matrix has been
proposed. Classical examples are KD encoding methods 
\cite{chen2018learning,shu2017compressing}. It uses $k$-way and $d$-dimensional
discrete encoding scheme to replace the embedding vector. 
However, a specialized method is required to train the KD codes, and the set of
symbols has to be fixed. In contrast, the set of features in our
problem varies as training proceeds and the lengths of embedding vectors are
different.

Model compression is a traditional method to reduce model size.
Parameter sharing, first proposed as HashedNet in \cite{chen2015compressing}, has
been used to reduce the memory
footprint of over-parameterized deep learning models. 
HashedNet groups the neural connections into hash
buckets uniformly at random via hash functions before the model sees
any data, i.e. the sharing does not use any training information. 
Deep compression \cite{han2015deep} also proposes a parameter sharing method. 
It clusters the well-trained parameters and retrains the model to compensate for
the
accuracy loss. 
In these cases, the number of parameters is fixed and it is difficult to apply
them to embedding matrix with a varying number of parameters as we explained in
 \cref{sec:intro}. 
The processing unit of parameter sharing is weight,
which is different from the embedding vectors. 

Another closely related work is \cite{covington2016deep} where the authors
use only the high frequency features for Youtube recommendation system.
Since low frequency features cannot be trained well, they are simply assigned
with the zero embedding. In contrast, \sys groups the low frequency features
with similar high frequency features and performs retraining for better results.

\section{Conclusion}
We have presented \sys, a novel method to compress the embedding matrix in
recommendation systems. 
\sys clusters the most frequent features based on their similarity to reduce the
number of embedding vectors and accelerate the clustering speed. 
\sys assigns the low frequency features to their closest centroids which
can alleviate the effect of noise and insufficient training. 
Testbed experiments based on implementation in \xr{a production
system} shows that \sys achieves similar or better
prediction performance compared to the original model with about 27x effective 
compression ratio. 

\bibliographystyle{abbrv}
\small
\bibliography{ms}

\end{document}